\documentclass[12pt]{iopart}

\usepackage{graphicx}
\usepackage{iopams}
\usepackage{caption}
\newif\ifarxiv
\arxivtrue
\begin{document}

\title[Elliptic Flow Fluctuations by PHOBOS]{Non-flow correlations and
elliptic flow fluctuations in Au+Au collisions at $\sqrt{s_{_{\it
NN}}} =$~200~GeV} \author{B Alver for the PHOBOS collaboration \ifarxiv \else \footnote{For the full list of PHOBOS authors and acknowledgments, see appendix “Collaborations”.} \fi}
\address{%Laboratory for Nuclear Science,
Massachusetts Institute of Technology, %77 Massachusetts Ave.  
Cambridge, MA 02139, USA}
\ead{alver@mit.edu}

\ifarxiv 
\author{
B.Alver$^4$,
B.B.Back$^1$,
M.D.Baker$^2$,
M.Ballintijn$^4$,
D.S.Barton$^2$,
R.R.Betts$^6$,
A.A.Bickley$^7$,
R.Bindel$^7$,
W.Busza$^4$,
A.Carroll$^2$,
Z.Chai$^2$,
V.Chetluru$^6$,
M.P.Decowski$^4$,
E.Garc\'{\i}a$^6$,
T.Gburek$^3$,
N.George$^2$,
K.Gulbrandsen$^4$,
C.Halliwell$^6$,
J.Hamblen$^8$,
I.Harnarine$^6$,
M.Hauer$^2$,
C.Henderson$^4$,
D.J.Hofman$^6$,
R.S.Hollis$^6$,
R.Ho\l y\'{n}ski$^3$,
B.Holzman$^2$,
A.Iordanova$^6$,
E.Johnson$^8$,
J.L.Kane$^4$,
N.Khan$^8$,
P.Kulinich$^4$,
C.M.Kuo$^5$,
W.Li$^4$,
W.T.Lin$^5$,
C.Loizides$^4$,
S.Manly$^8$,
A.C.Mignerey$^7$,
R.Nouicer$^2$,
A.Olszewski$^3$,
R.Pak$^2$,
C.Reed$^4$,
E.Richardson$^7$,
C.Roland$^4$,
G.Roland$^4$,
J.Sagerer$^6$,
H.Seals$^2$,
I.Sedykh$^2$,
C.E.Smith$^6$,
M.A.Stankiewicz$^2$,
P.Steinberg$^2$,
G.S.F.Stephans$^4$,
A.Sukhanov$^2$,
A.Szostak$^2$,
M.B.Tonjes$^7$,
A.Trzupek$^3$,
C.Vale$^4$,
G.J.van~Nieuwenhuizen$^4$,
S.S.Vaurynovich$^4$,
R.Verdier$^4$,
G.I.Veres$^4$,
P.Walters$^8$,
E.Wenger$^4$,
D.Willhelm$^7$,
F.L.H.Wolfs$^8$,
B.Wosiek$^3$,
K.Wo\'{z}niak$^3$,
S.Wyngaardt$^2$,
B.Wys\l ouch$^4$\\
}
\vspace{3mm}
\small
\address{ $^1$Argonne National Laboratory, Argonne, IL 60439, USA\\
$^2$Brookhaven National Laboratory, Upton, NY 11973, USA\\
$^3$Institute of Nuclear Physics PAN, Krakow, Poland\\
$^4$Massachusetts Institute of Technology, Cambridge, MA 02139, USA\\
$^5$National Central University, Chung-Li, Taiwan\\ $^6$University of
Illinois at Chicago, Chicago, IL 60607, USA \\ $^7$University of
Maryland, College Park, MD 20742, USA\\ $^8$University of
Rochester,Rochester, NY 14627, USA} \fi

\begin{abstract} We present first results on event-by-event elliptic
flow fluctuations in nucleus-nucleus collisions corrected for effects
of non-flow correlations where the magnitude of non-flow correlations
has been independently measured in data. Over the measured range in
centrality, we see large relative fluctuations of 25-50\%.  The
results are consistent with predictions from both color glass
condensate and Glauber type initial condition calculations of the
event-by-event participant eccentricity fluctuations.
\end{abstract}

\section{Introduction} Elliptic flow is one of the key observables in
understanding the early stages and the dynamics of heavy ion
collisions. Large elliptic flow signals have been observed at the top
RHIC energies in both Au+Au and Cu+Cu
collisions~\cite{firstpaper,auaupaper,cucupaper}. These results have
been understood in terms of the initial anisotropy in the collision
region, best described by the participant
eccentricity~\cite{cucupaper,eccpaper}, being preserved via an early
thermalisation after the collision and the hydrodynamic expansion of
the system with very little viscosity~\cite{Teaney2003}. Initial
measurement of $v_2$ fluctuations appear to confirm this picture where
the event-by-event fluctuations in the initial shape anisotropy,
quantified by the participant eccentricity are translated to
event-by-event fluctuations in the $v_2$ signal~\cite{mypaper}.

Particle correlations other than flow~(non-flow correlations) such as
HBT, resonance decays and jets can resemble correlations due to
elliptic flow and can have various effects on different flow
measurements. In particular, non-flow correlations can broaden the
observed event-by-event $v_2$ distribution and could be mistaken for
$v_2$ fluctuations. Quantitative understanding of the effect of
non-flow correlations on the observed $v_2$ fluctuations requires a
measurement where the flow and non-flow correlations are completely
disentangled. In this paper, we introduce a new method of measuring
the non-flow correlations which crucially relies on the large
pseudo-rapidity coverage of the PHOBOS octagon multiplicity detector.
We present results on non-flow correlation strength and elliptic flow
fluctuations corrected for non-flow correlations in Au+Au collisions
at $\sqrt{s_{_{\it NN}}} =$~200~GeV.

\section{Measurement of non-flow correlations} We aim to measure the
second Fourier coefficient of two-particle angular correlations and
separate contributions from flow and non-flow in this signal. Flow and
non-flow contributions can be separated with a detailed study of the
$\eta$ and $\Delta\eta$ dependence of the $\Delta\phi$ correlation
function. To this end we define the two particle correlation function
between particles from two $\eta$ windows centered at $\eta_1$ and
$\eta_2$:
\begin{equation}
R_n(\Delta\phi,\eta_1,\eta_2)=\frac{F(\Delta\phi,\eta_1,\eta_2)}{B(\Delta\phi,\eta_1,\eta_2)}-1,
\end{equation} where $F$ is the foreground distribution determined by
taking hit pairs from the same event and $B$ is the background
distribution constructed by randomly selecting particles from two
different events with similar vertex position and centrality. In
practice, the correlation function is calculated using the procedure
described in Ref.~\cite{weispaper}, correcting for the effect of the
particles produced due to the interactions with the detector.

It can be shown that if the only correlations between particles are
due to elliptic flow, the correlation function will take the form:
$R_n(\Delta\phi,\eta_1,\eta_2)=2v_2(\eta_1)\times
v_2(\eta_2)\cos(2\Delta\phi)$. In general, both flow and non-flow effects are present and so we denote the second Fourier coefficient
of $R_n(\Delta\phi,\eta_1,\eta_2)$ as $2v_2^2(\eta_1,\eta_2)$.  The seperate flow and non-flow contributions are denoted $v_2(\eta)$ and $\delta(\eta_1,\eta_2)$
respectively, such that:
\begin{equation} v_2^2(\eta_1,\eta_2)=v_2(\eta_1)\times
v_2(\eta_2)+\delta(\eta_1,\eta_2)
\end{equation} In Fig.~\ref{fig:v2sqr}, $v_2^2(\eta_1,\eta_2)$ is
shown for a selected centrality range in Au+Au collisions at
$\sqrt{s_{_{\it NN}}} =$~200~GeV. The contribution of two different
sources can be observed: a flow plateau, $v_2(\eta_1)\times
v_2(\eta_2)$, which is seperable in $\eta_1$ and $\eta_2$ and a
non-flow ridge along the diagonal $\eta_1=\eta_2$.

At large pseudo-rapidity separations, i.e.\
$\Delta\eta\!\equiv\!\left|\eta_1-\eta_2\right|>2$, we expect the
non-flow component $\delta(\eta_1,\eta_2)$ to be small and estimate it
by a comparison of the Fourier expansion of the $\Delta\phi$
correlations between data and HIJING. From these studies we estimate
roughly that the non-flow in data in this region is in the range 0 to 3.2
times the non-flow magnitude in HIJING. The uncertainty in the
estimation of $\delta(\eta_1,\eta_2)$ for $\Delta\eta>2$ is the
dominant source of error in the final magnitude of non-flow
correlations. Further studies are planned to improve the understanding
of the non-flow magnitude in this region.

Performing a separable fit of $v_2(\eta_1)\times v_2(\eta_2)$ to
$v_2^2(\eta_1,\eta_2)-\delta(\eta_1,\eta_2)$ in the region
$\Delta\eta>2$, where $v_2^2(\eta_1,\eta_2)$ is measured in data and
$\delta(\eta_1,\eta_2)$ is taken as $1.6\pm1.6$ times the magnitude in
HIJING, we can measure $v_2(\eta)$. The fit in the selected
$\Delta\eta$ region can be used to extract the correlation magnitude
due to flow, $v_2(\eta_1)\times v_2(\eta_2)$, in the whole
pseudorapidity accceptance $\left|\eta_1\right|<3$ and
$\left|\eta_2\right|<3$. Subtracting the correlations due to flow from
$v_2^2(\eta_1,\eta_2)$ we can calculate $\delta(\eta_1,\eta_2)$ in all
regions of $\eta_1$ and $\eta_2$.

\section{Elliptic flow fluctuations corrected for non-flow
correlations} 
\begin{figure}[t]
  \begin{minipage}[t]{0.27\linewidth}
    \includegraphics[width=0.97\textwidth]{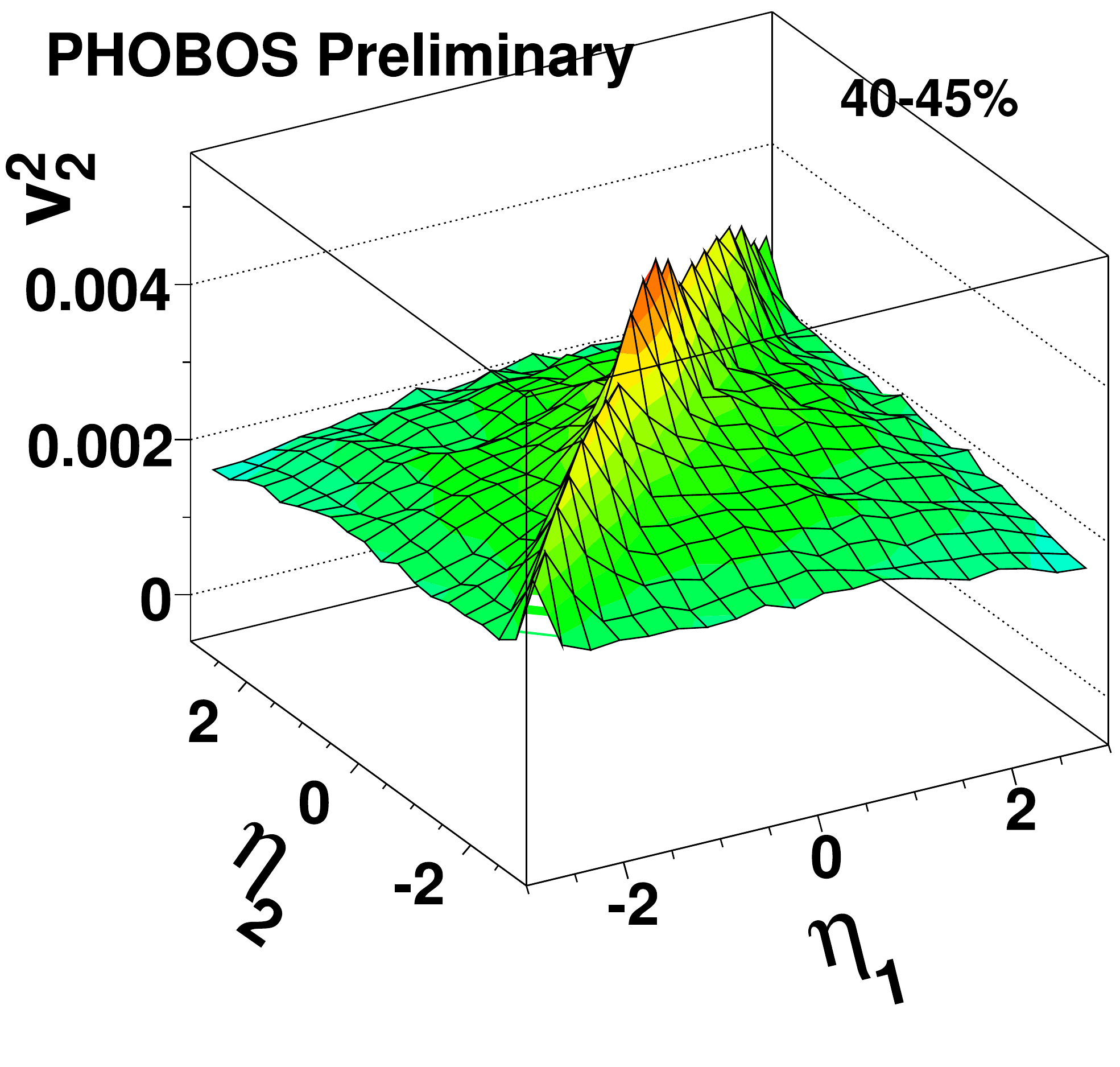}
    \caption{Second Fourier coefficient of the correlation function
$R_n(\Delta\phi,\eta_1,\eta_2)$ as a function of $\eta_1$ and $\eta_2$
for the 40-45\% central Au+Au collisions at $\sqrt{s_{_{\it NN}}}
=$~200~GeV.}
    \label{fig:v2sqr}
  \end{minipage} \hspace{0.01\linewidth}
  \begin{minipage}[t]{0.32\linewidth}
    \includegraphics[width=0.97\textwidth]{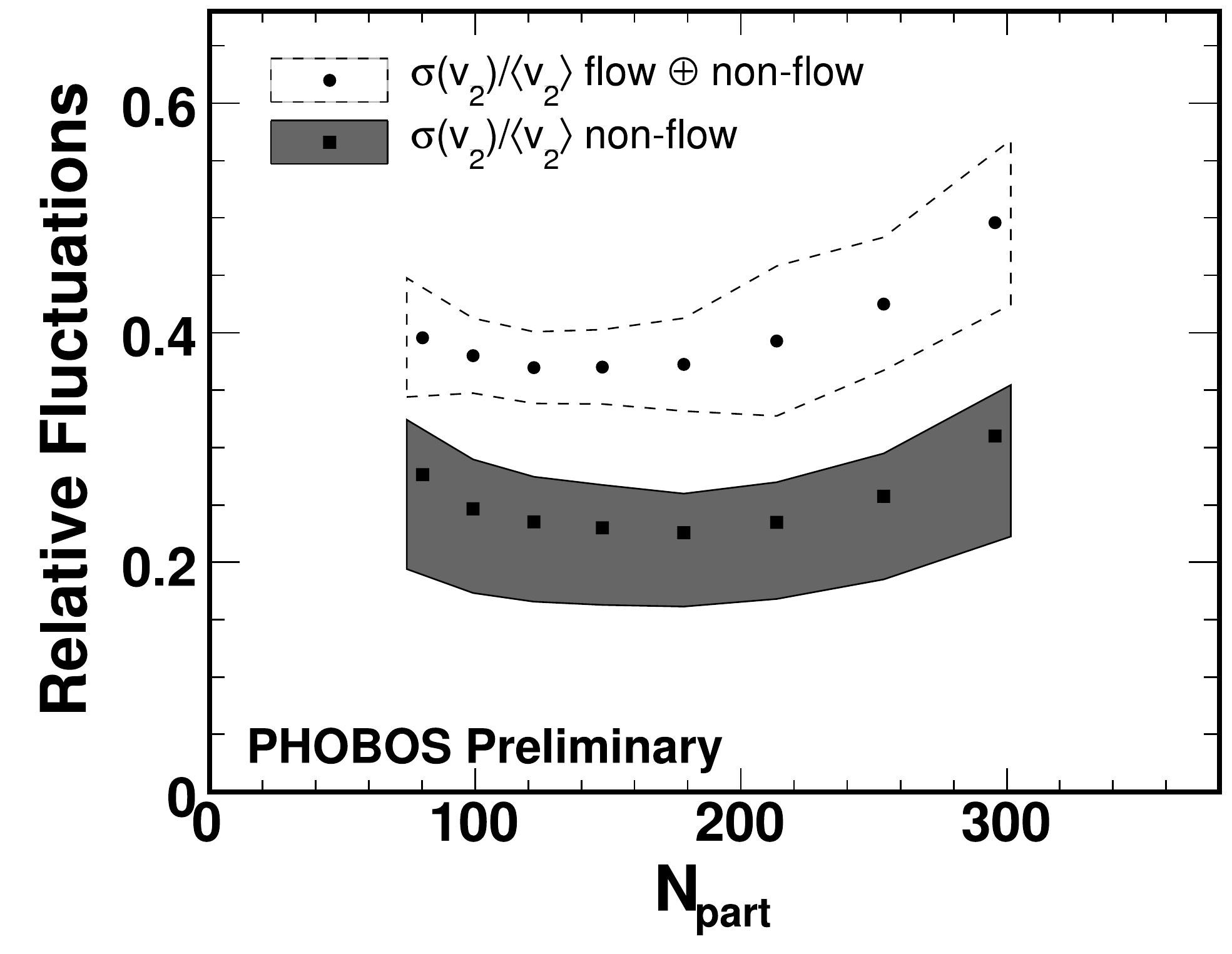}
    \caption{Circles:
Total measured fluctuations including flow fluctuations and
effects of non-flow correlations~\cite{mypaper}. Squares: Expected
measured fluctuations for the observed non-flow correlation
signal.}
    \label{fig:totdel}
  \end{minipage} \hspace{0.01\linewidth}
  \begin{minipage}[t]{0.32\linewidth}
    \includegraphics[width=0.97\textwidth]{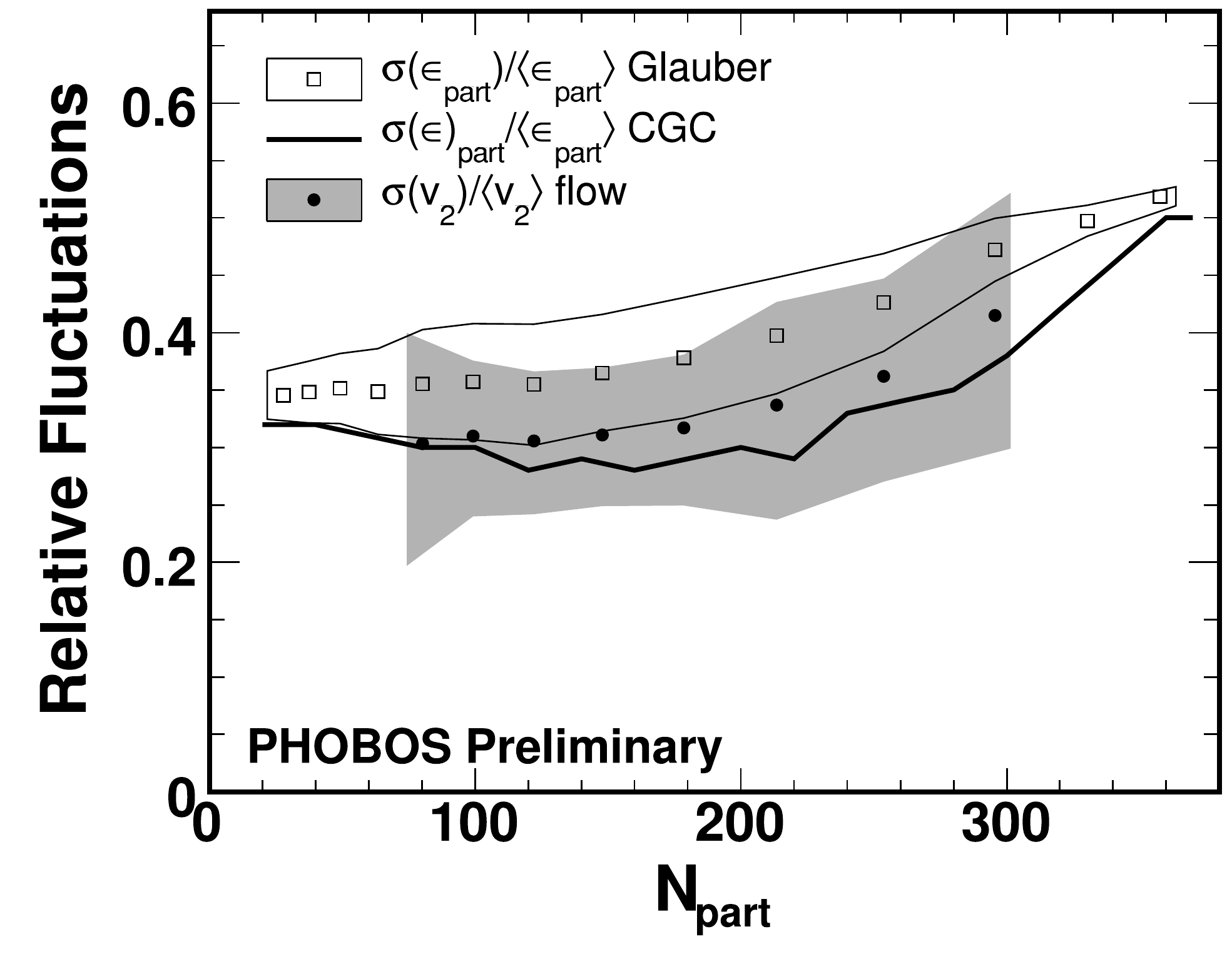}
    \caption{Circles:
Results corrected for effects of non-flow
correlations. Thick black line ands squares show
$\sigma(\epsilon_{part})/\langle\epsilon_{part}\rangle$ calculated in
CGC~\cite{cgc} and Glauber MC~\cite{mypaper} models,
respectively.}
    \label{fig:flowecc}
  \end{minipage} \vspace{-0.3cm}
\end{figure}

If there are no true flow fluctuations, the measurement
of event-by-event $v_2$ fluctuations will yield an RMS width
of $\sigma(v_2)_{\textrm{non-flow}}=\sqrt{\langle \delta \rangle /2}$,
where $\langle\delta\rangle$ is the average $\cos(2\Delta\phi)$ due to
non-flow correlations over all particle pairs~\cite{gaussian}. Since the
second Fourier coefficient of $R_n(\Delta\phi,\eta_1,\eta_2)$ is equal
to $2\langle\cos(2\Delta\phi)\rangle$, the $v_2^2$ defined above equals $\langle\cos(2\Delta\phi)\rangle$. Therefore we can calculate $\langle\delta\rangle$ by averaging $\delta(\eta_1,\eta_2)$, the non-flow component of $v_2^2$ for particles in the two windows centered at $\eta_1$ and $\eta_2$. Averaging over all particle pairs gives:
\begin{equation} \langle \delta
\rangle=\frac{\int\delta(\eta_1,\eta_2)\frac{{\rm d}N}{{\rm d}\eta_1}
\frac{{\rm d}N}{{\rm d}\eta_2} {\rm d}\eta_1 {\rm
d}\eta_2}{\int\frac{{\rm d}N}{{\rm d}\eta_1} \frac{{\rm d}N}{{\rm
d}\eta_2} {\rm d}\eta_1 {\rm d}\eta_2}
\end{equation} The expected relative fluctuations due to non-flow,
$\sigma(v_2)_{\textrm{non-flow}}/\langle v_2 \rangle$ is shown in
comparison to the the total measured relative $v_2$ fluctuations in
Fig.~\ref{fig:totdel}.

When we study the response of the $v_2$ fluctuations analysis on event
samples with flow fluctuations and non-flow correlations, we find the
following empirical dependence:
\begin{equation} \sigma(v_2)_{\textrm{total}} =
\sigma(v_2)_{\textrm{flow}} + \sigma(v_2)_{\textrm{non-flow}} \times
\exp
\left(-\frac{\sigma(v_2)_{\textrm{flow}}}{\sigma(v_2)_{\textrm{non-flow}}}\right),
  \label{eq:empir}
\end{equation} where $\sigma(v_2)_{\textrm{total}}$ refers to the
measured $v_2$ fluctuations and $\sigma(v_2)_{\textrm{flow}}$ refers
to the true flow fluctuations. Subtracting the expected fluctuations
due to non-flow correlations, $\sqrt{\langle \delta \rangle /2}$, from
the measured total $v_2$ fluctuations via Eq.~\ref{eq:empir}, we
extract the corrected elliptic flow fluctuations, shown in
Fig.~\ref{fig:flowecc}.

Also shown in Fig.~\ref{fig:flowecc} are $\sigma_{\epsilon_{{\rm
part}}}/\langle \epsilon_{{\rm part}} \rangle$ at fixed values of
$N_{part}$ obtained in MC Glauber~\cite{mypaper} and color glass
condensate(CGC)~\cite{cgc} calculations.  The 90\% confidence level
systematic errors for MC Glauber calculations (shown as a band in Fig.~\ref{fig:flowecc}) are estimated by varying
Glauber parameters as discussed in Ref.~\cite{cucupaper}. The relative
flow fluctuations are in agreement with participant eccentricity
predictions calculated both with MC Glauber and CGC
type initial conditions over the full centrality range under study.
The observed agreement suggests that the fluctuations of elliptic flow
primarily reflect fluctuations in the initial state geometry and are
not affected strongly by the latter stages of the collision.

These results are therefore qualitatively consistent with a picture of
the collision process in which the shape of the initial stage geometry
follows the predictions of the participant eccentricity model and
where the initial geometry is translated into the final state
azimuthal particle distribution in a hydrodynamic expansion, leading
to an event-by-event proportionality between the observed elliptic
flow and the initial eccentricity. 
The results, however, do not allow us to distinguish between
Glauber and color glass condensate type initial conditions.

\section{Conclusion}

In summary, we have presented the first measurement of elliptic flow
fluctuations corrected for non-flow effects extracted independently
from data in Au+Au collisions at $\sqrt{s_{_{\it NN}}} =$~200~GeV. A
significant non-flow correlation strength is observed. However, even
after correcting for these correlations, the elliptic flow fluctuation
signal is seen to be large with a relative magnitude of ~25-50\%. We
show that the magnitude and centrality dependence of these
fluctuations are in agreement with predictions for fluctuations of the
initial shape of the collision region quantified by the participant
eccentricity calculated with either Glauber or color glass condensate
type initial conditions. 
These results support conclusions from previous studies
on the importance of geometric fluctuations of the initial collision
region postulated to relate elliptic flow measurements in the Cu+Cu
and Au+Au systems~\cite{cucupaper}. The initial geometry seems to
drive the hydrodynamic evolution of the system, not only on average,
but event-by-event.

\ifarxiv
This work was partially supported by U.S. DOE grants 
DE-AC02-98CH10886,
DE-FG02-93ER40802, 
DE-FG02-94ER40818,  % MIT
DE-FG02-94ER40865, 
DE-FG02-99ER41099, and
DE-AC02-06CH11357, by U.S. 
NSF grants 9603486, % Phobos TOF 
0072204,            % Rochester until 6/03
and 0245011,        % Rochester starting 6/03
by Polish MNiSW grant N N202 282234 (2008-2010),
by NSC of Taiwan Contract NSC 89-2112-M-008-024, and
by Hungarian OTKA grant (F 049823).
\fi

\end{document}